\documentclass[preprint,3p]{elsarticle}
\usepackage{subcaption}
\usepackage{graphicx}
\usepackage{amsthm}
\usepackage{verbatim}
\usepackage{dcolumn}
\usepackage{bm}
\usepackage{epsf}
\usepackage{color}
\usepackage[colorlinks=true,citecolor=blue,linkcolor=blue,urlcolor=blue]{hyperref}%
\biboptions{sort&compress}
\usepackage{xcolor}
\usepackage{dsfont}
\usepackage{tikz}
\usetikzlibrary{decorations.pathreplacing,decorations.pathmorphing}

\newcommand{\bla}{bla\\bla\\bla\\bla\\bla}

\makeatletter
\newcommand{\currentfontsize}{\f@size pt}
\makeatother
\newcommand\figsize{1}

\begin{document}

\begin{frontmatter}

\title{Simon's algorithm in the NISQ cloud}
\author[1,2]{Reece Robertson \corref{cor1}}
\ead{reecerobertson@umbc.edu}
\author[2]{Emery Doucet}
\author[3]{Ernest Spicer}
\author[2,4]{Sebastian Deffner}
\ead[url]{quthermo.umbc.edu}
\affiliation[1]{organization={Department of Computer Science and Electrical Engineering, University of Maryland, Baltimore County}, 
city={Baltimore}, 
postcode={MD 21250},
country={USA}}
\affiliation[2]{organization={Department of Physics, University of Maryland, Baltimore County}, 
city={Baltimore}, 
postcode={MD 21250},
country={USA}}
\affiliation[3]{organization={sagax.ai}, 
city={Pullman}, 
postcode={WA 99163},
country={USA}}
\affiliation[4]{
organization={National Quantum Laboratory}, 
city={College Park}, 
postcode={MD 20740},
country={USA}}

\cortext[cor1]{Corresponding author}
    

\begin{abstract}
Simon’s algorithm was one of the first problems to demonstrate a genuine quantum advantage.
The algorithm, however, assumes access to noise-free qubits.
In our work we use Simon’s algorithm to benchmark the error rates of devices currently available in the ``quantum cloud.''
As a main result we obtain an objective comparison between the different physical platforms made available by IBM and IonQ.
Our study highlights the importance of understanding the device architectures and chip topologies when transpiling quantum algorithms onto hardware.
For instance, we demonstrate that two-qubit operations on spatially separated qubits on superconducting chips should be avoided.
\end{abstract}


\end{frontmatter}

\section{Introduction}

Recent years have seen the rapid increase in investments into quantum technologies.
Some market analysts even project the global volume to gain up to  US\$1.3 trillion in value by the mid 2030s \cite{invest}.
This unprecedented growth in economic interest is driven by fact that fully functioning quantum devices will have the ability to exponentially outperform classical technologies.
This \emph{quantum advantage} originates in the peculiarities of quantum physics, which allows quantum technologies to perform certain tasks with exponentially less resources than would be required by a classical device \cite{bible,Sanders2017,Savage2017}.

Global excitement has been further fueled by the first couple of experiments demonstrating \emph{quantum advantage} for judiciously designed computational tasks \cite{Arute2019,HanSen2020Science,Wu2021PRL,Zhong2021PRL,Madsen2022Nature,Zhu2022SB,King2024arXiv}.
However, currently available hardware is somewhat ``small'' and, more importantly, it is still prone to noise.
Despite continuing progress in the development of fault-tolerant qubits \cite{resilient}, there is general consensus that we are currently in the era of Noisy, Intermediate-Scale Quantum (NISQ) devices \cite{Preskill2018quantum,taxonomy,nisq_computing,survey,evaluation,benchmarking,stability,characterizing,gate-based,supermarq,qpack,qasmbench,nisq_analyzer}.
This raises the natural question of how these NISQ devices perform when tested against genuine quantum algorithms that were designed for the very purpose to be quantum advantageous.

Arguably the most famous algorithm is due to Shor \cite{shor}, which was the first to demonstrate potential applications of quantum computing to ``practical problems''---namely factorization into prime numbers.
Remarkably, the first experimental implementations were already reported almost three decades ago \cite{chuang} albeit at very small scales.
Similarly, experiments investigated the performance random circuits \cite{frontier}, Grover's search  and Bernstein-Vazirani algorithms \cite{nisq_complexity}.
Importantly, it was concluded that fair sampling has not been achieved on any of the devices by IBM, Rigetti, IonQ, and DWave \cite{sampling}.

In the present paper, we use Simon's algorithm \cite{simon} to benchmark several NISQ devices available for remote, cloud access.
This algorithm is significant since it was among the first to use the quantum Turning machine framework to obtain a provable exponential speedup over its classical probabilistic counterpart \cite{bible}.
In other words, it was among the first to show that a quantum Turing machine can violate the Church-Turing thesis \cite{strengths_weaknesses}.
Interestingly, the resilience of Simon's algorithm was tested by performing Monte Carlo simulations of stochastic Pauli noise operators \cite{impact}.
It was concluded that Simon's algorithm is rather susceptible to imperfections and noise, and hence the performance of the algorithm makes for a very sensitive tool for the noise diagnostics of NISQ hardware.

We implemented two versions of Simon's algorithm, with minimally and maximally required entangling operations; the instances were run on three versions of a superconducting platform available at IBM, and on three versions of an ion-trapped device by IonQ.
As a measure of performance, we computed the percentage of iterations that returned an invalid answer at each problem size; a metric we refer to as the \textit{algorithmic error rate}.
For each physical platform we compared the resulting error rate with the prediction of the noisy simulators provided by the quantum computing companies.

As main results we found that all NISQ computers examined experience an increase in algorithmic error as the problem size increase.
This is an important observation as all devices exhibit an (extrapolated) error rate of ~50\% for problems requiring more than 50 qubits\footnote{Depending on the details of the estimate, more than about 53 qubits are required for the unambiguous emergence of quantum advantage \cite{Arute2019}.}.
This means that for problems large enough to possibly be quantum advantageous the algorithm completely fails (as an error rate of ~50\% corresponds to random guessing).
It is even more striking to observe that none of the noisy simulators quantitatively capture the behavior of the physical devices.
Moreover, we found that the error rate scales linearly for the devices of IonQ, and that for the superconducting platforms of IBM stark departure from linear scaling is present for the most complex algorithm.
We trace this observation back to the entangling gates between spatially separated qubits.
Hence, we draw two main conclusions, namely (i) NISQ devices available for remote access are still too noisy to support genuine quantum advantage and (ii) when transpiling algorithms to physical platforms close attention needs to be paid to the QPU architecture.

\section{Hidden subgroups and Simon's problem}
\label{background}

A good starting point for the quest for quantum advantage are \emph{hidden subgroup problems} \cite{bible}.
In simple terms, such problems refer to decrypting information that is ``hidden'' from direct access.
Typical examples include factoring, evaluating the discrete logarithm and graph isomorphisms, and the shortest vector problem.
Arguably the most prominent example of hidden subgroup problems is Shor's  algorithm \cite{Shor1994}, which in turn was inspired by Simon's problem \cite{simon}.

As it was originally posed, Simon's algorithm distinguishes between two classes of functions that operate on bitstrings of size $n$.
Assume that we are given a function $f$, which is either one-to-one, or two-to-one with the property that there exists some ``secret string'' $s$ such that for every two inputs that map to the same output, the \texttt{XOR} of the inputs is $s$ \cite{simon, mermin}.
In other words, $f$ is either a one-to-one function or it is a two-to-one periodic function with period $s$.
Further assume that we have access to a black-box quantum operation, an \emph{oracle}, $U_f$ that operates on two quantum registers $|x\rangle$ and $|y\rangle$.
Then, $U_f$ evaluates $f$ on the contents of the first register and saves it in the second register. In mathematical terms,
\begin{equation}
U_f\left(|x\rangle|y\rangle\right) = |x\rangle|f(x) \oplus y\rangle\,,
\end{equation}
where $\oplus$ denotes addition modulo 2.

Given some $U_f$, Simon's problem is to classify the associated $f$ into one of the two classes \cite{simon}.
If $f$ is of the second class then the period $s$ of $f$ must also be identified.
Note that for the present work, we choose the provided function $f$ to be of the second class, that is, $f$ is two-to-one with period $s$.

\begin{figure}
    \centering
    \includegraphics[width=\textwidth]{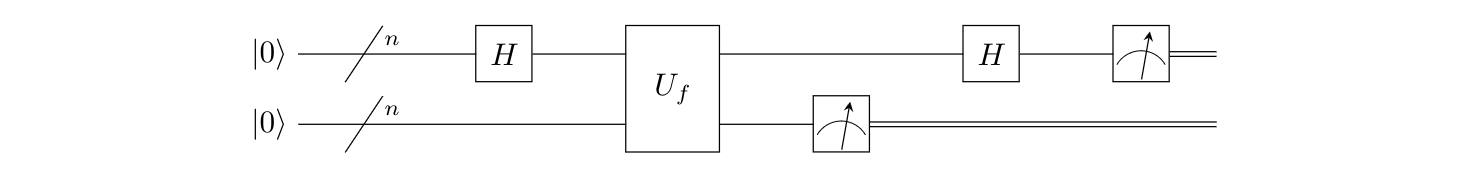}
    \caption{
        Quantum circuit diagram for Simon's algorithm:
        Note the use of two quantum registers of size $n$, both initialized to the zero state $|0\rangle$, as well as the oracle $U_f$.
        The algorithm uses two Hadamard transformations and $U_f$ to create a superposition over all the size $n$ bitstrings that are orthogonal to the secret string $s$.
        An iteration of the algorithm is successful if the final measurement result is indeed orthogonal to $s$ (this is always true in the noise free case).
        The entire algorithm is successful if a complete set of $n-1$ linearly independent bitstrings are measured and the resulting system is solved classically in polynomial-time.
    }
    \label{fig:circuit}
\end{figure}

For clarity, we depict the corresponding quantum circuit in Fig.~\ref{fig:circuit}.
For the following analysis it will be instructive to summarize the explicit steps of the implementation:
Simon's algorithm prepares two quantum registers each containing $n$ qubits in the ground state $|0\rangle$, before a Hadamard transformation is applied to the first register.
Then $U_f$ is applied, which stores the evaluation of $f$ on all possible inputs in the second quantum register.
The second register may then be measured to select a single element of the image of $f$.
If this is done then the superposition of the entangled first register correspondingly collapses to a superposition of the two elements of the domain of $f$ which map to this output\footnote{This last step is entirely optional but it simplifies the mathematics of the problem and is often included in presentations of the algorithm}.
A second Hadamard transformation is then applied to the first register, and after that it is measured as well.

This second Hadamard transformation is a critical step of the algorithm.
It creates an interference pattern on the first register resulting in the useful superposition of all elements of the domain of $f$ which are orthogonal to the secret string $s$.
Finally, the first register is measured.
If the algorithm has worked properly then this measurement will always return a bitstring that is orthogonal to the secret string $s$.
Repeating the quantum circuit gives a different bitstring that is also orthogonal to $s$ with high probability.
With $n-1$ linearly independent bitstrings orthogonal to $s$, we can construct a system of equations with sufficient information to solve for $s$ in polynomial-time using a classical linear solver.

In the following, we report the performance of this implementation of Simon's algorithm on NISQ devices.
This is particularly interesting, since Simon's algorithm was one of the first to provide a provable divergence in runtime between a probabilistic Turing machine and a quantum Turing machine.
Solving this problem scales exponentially on probabilistic, classical machines, while scaling polynomially on quantum computers \cite{simon}.
However, this speedup is predicated on the assumption that one has access to a sufficient number of \emph{noise-free} qubits.
When running Simon's algorithm on real quantum devices it is to be expected that this is not the case.
Instead, some percentage of the iterations of an instance of Simon's algorithm will produce results that are \textit{not} orthogonal to $x$.
The main objective of our analysis is to investigate how poorly the algorithm performs on real hardware, and what we can learn about the origins of the failure from the results.

\section{The quantum cloud}

Over the last decade, several large corporation as well as smaller start-up companies have made their NISQ devices available for cloud access.
This offers exciting prospects for fundamental as well as applied research, as now with comparatively little effort ``quantum experiments'' can be performed.
However, it has also already been demonstrated that the reported \emph{quantum volume}\footnote{The \emph{quantum volume} is defined using the number of qubits and the number of operations that can be effectively implemented on the device \cite{Cross2019PRA}.}
often overestimates was is experimentally accessible \cite{quantum_volume}.

This poses the somewhat natural question, how do algorithms designed for the very purpose of exhibiting quantum advantage perform on available NISQ devices?
For our study, we had access to IBM's superconducting platform and IonQ's trapped ion device.

\subsection{Superconducting qubits -- IBM}

The web application called \emph{IBM Quantum}, formerly known as \emph{IBM Quantum Experience}, provides cloud-based access to a variety of tutorials, simulators, and real quantum processors.
The service was originally launched in May 2016, and over the last decade the systems have gained some maturity.

IBM's quantum processors are made up of superconducting transmon qubits \cite{Koch2007PRA}, located in dilution refrigerators at the IBM Research headquarters at the Thomas J. Watson Research Center.
The original system consisted of only 5 qubits connected in a star geometry, which, however, already supported studies of fundamental questions in physics, see e.g. Ref.~\cite{Deffner2017Heliyon}.
The latest generation, the Condor chip has more than 1000 qubits \cite{Castelvecchi2023ibm}.
This has created a lot of interest in the quantum community, and for instance Ref.~\cite{cloud} has analyzed IBM cloud data (including job time, queue time, compilation time, etc.) and compared it to classical high performance computing cloud services.

\begin{figure}
    \centering
    \begin{subfigure}{.3\textwidth}
        \centering
        \includegraphics[width=\textwidth]{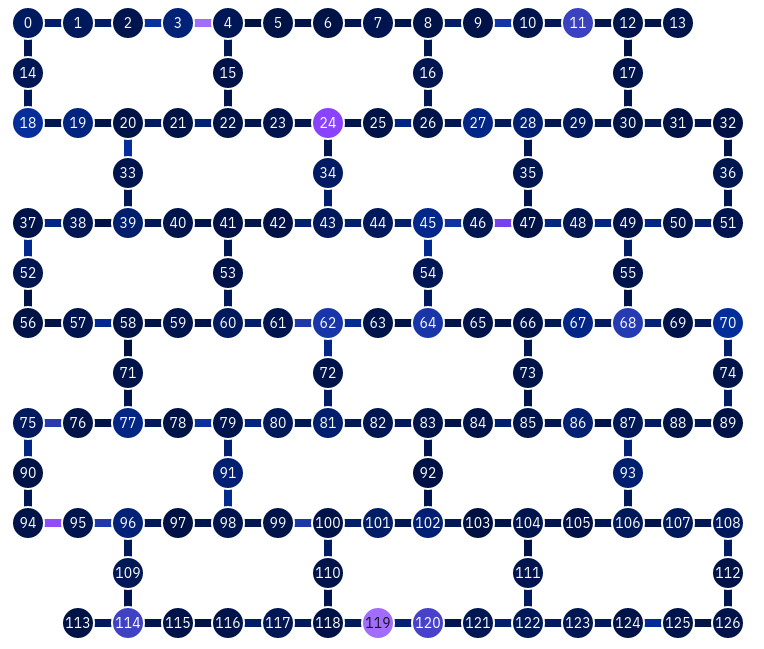}
        \caption{IBM Brisbane}
        \label{fig:brisbane-topology}
    \end{subfigure}
    \hfill
    \begin{subfigure}{.3\textwidth}
        \centering
        \includegraphics[width=\textwidth]{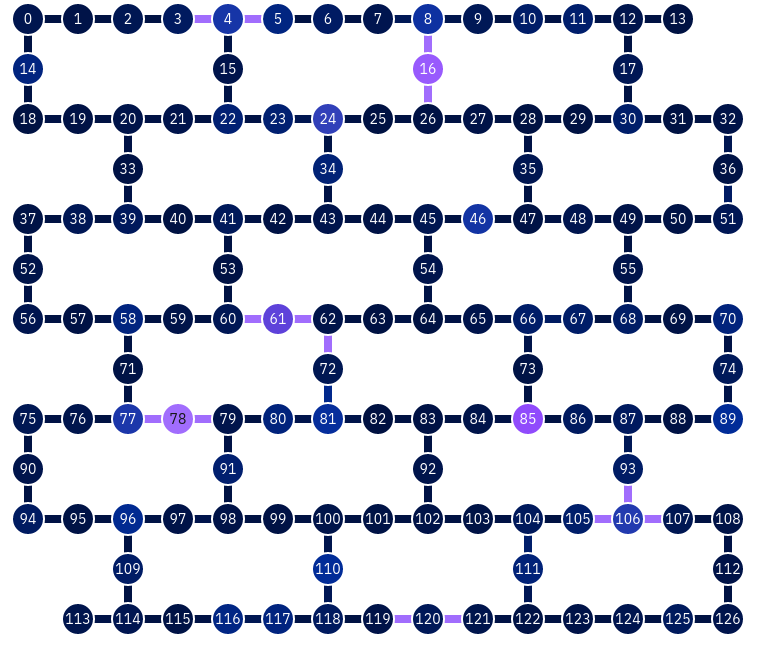}
        \caption{IBM Osaka}
        \label{fig:osaka-topology}
    \end{subfigure}
    \hfill
    \begin{subfigure}{.3\textwidth}
        \centering
        \includegraphics[width=\textwidth]{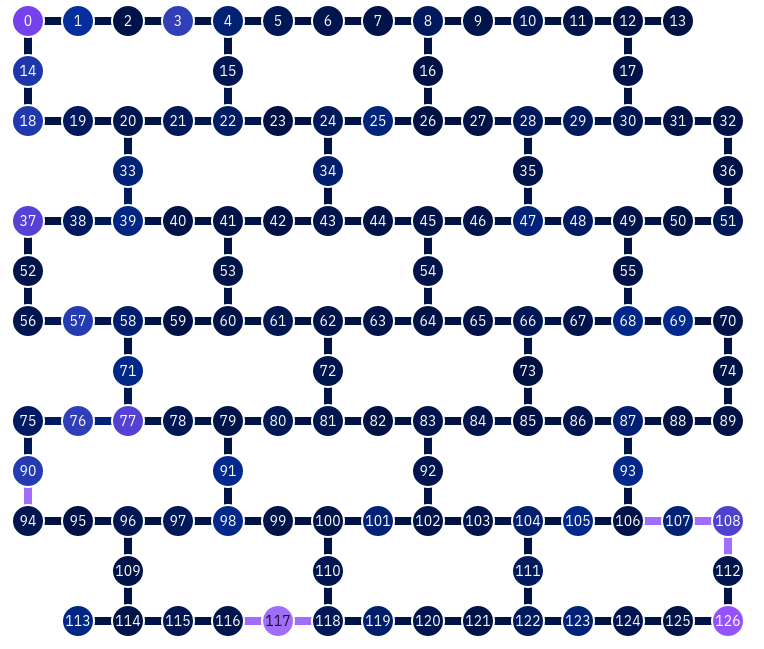}
        \caption{IBM Kyoto}
        \label{fig:kyoto-topology}
    \end{subfigure}
    \begin{subfigure}{\linewidth}
    \includegraphics[width=\textwidth]{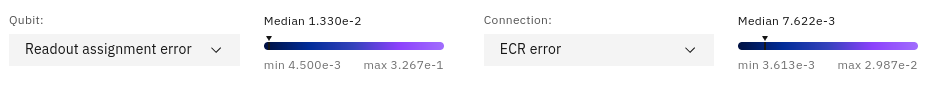}
    \label{fig:ibm-scale}
\end{subfigure}
    \caption{
        Device topologies for IBM computers:
        The color of qubits and connections represent the single qubit readout error and ECR error, respectively, at the time the image was taken.
        Lighter colors correspond to higher error rates.
        All images retrieved from \url{https://quantum.ibm.com/composer}.
    }
    \label{fig:ibm-topology}
\end{figure}

For our work, we used the 127-qubit Brisbane, Osaka, and Kyoto superconducting quantum processors, for which noisy simulators are also provided.
Quantum processing unit (QPU) topologies of each device are depicted in Fig.~\ref{fig:ibm-topology}.
Observe that all chip topologies are identical, as these three computers are all instances of the IBM Eagle chip design.

\subsection{Ion traps -- IonQ}

\begin{figure}
    \centering
    \begin{subfigure}{.3\textwidth}
        \centering
        \includegraphics[width=\textwidth]{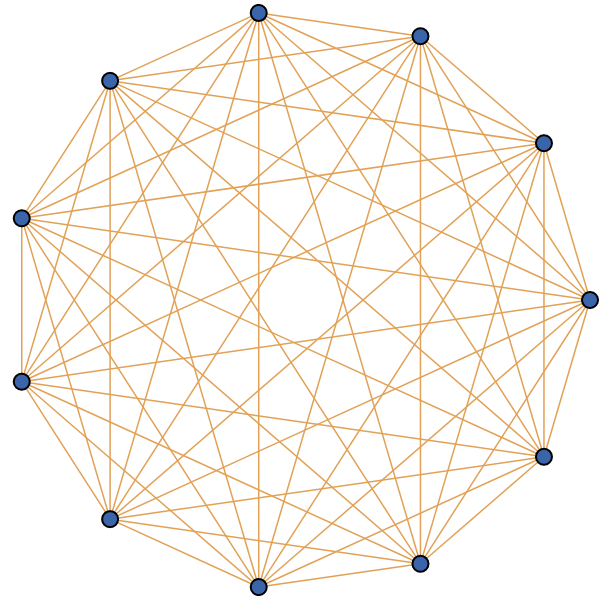}
        \caption{IonQ Harmony}
        \label{fig:harmony-topology}
    \end{subfigure}
    \hfill
    \begin{subfigure}{.3\textwidth}
        \centering
        \includegraphics[width=\textwidth]{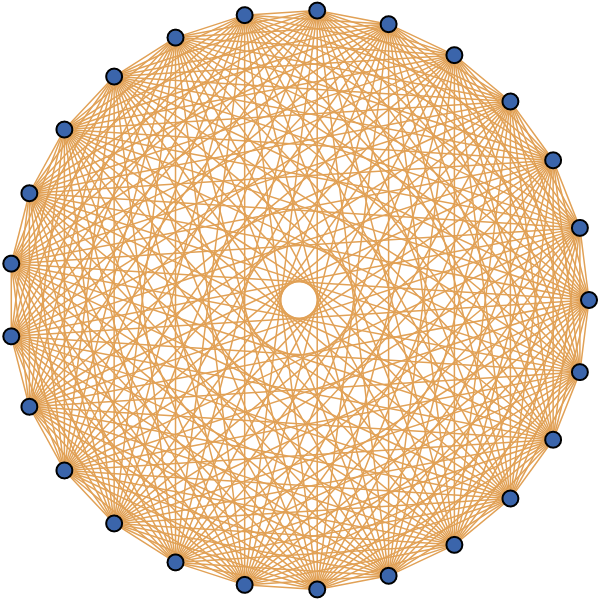}
        \caption{IonQ Aria 1}
        \label{fig:aria-topology}
    \end{subfigure}
    \hfill
    \begin{subfigure}{.3\textwidth}
        \centering
        \includegraphics[width=\textwidth]{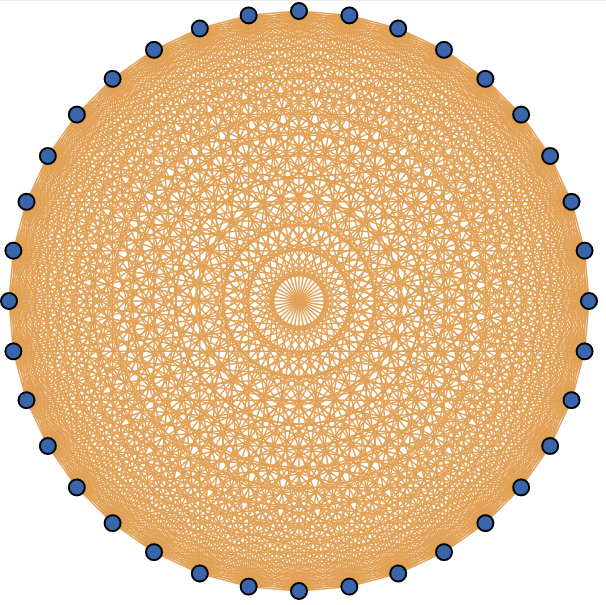}
        \caption{IonQ Forte 1}
        \label{fig:forte-topology}
    \end{subfigure}
    \caption{
        Device topologies for IBM computers:
        Observe that in trapped ion computers we have full all-to-all connectivity.
        All images retrieved from \url{https://us-west-2.console.aws.amazon.com/braket/home?region=us-west-2\#/devices}.
    }
    \label{fig:ionq-topology}
\end{figure}

A fundamentally different type of system are the trapped ion devices by IonQ \cite{Allen2017IEEE}, which are physically located in College Park, Maryland.
Typically, trapped ion NSIQ devices exhibit several advantages, such as accuracy, predictability, and coherence time \cite{Allen2017IEEE}.
However, such systems posses the disadvantage that the number of available qubits is much smaller.
The largest device of IonQ has ``only'' 36 qubits.
Some of the limitations arising from the small qubit number is made up by the fact that trapped ion computers poss full connectivity.
In Fig.~\ref{fig:ionq-topology} we depict the chip topologies of IonQ's Harmony, Aria, and Forte.

\section{Simon's algorithm on NISQ}
\label{approach}

\subsection{Implementation of the algorithm}

In the abstract formulation of Simon's algorithm, cf. Fig.~\ref{fig:circuit}, the oracle $U_f$ is an unspecified function.
This is sufficient for considerations of quantum advantage, and in fact the algorithm is formulate to yields results independent of the precise form of the oracle.

\begin{figure}
    \centering
    \begin{subfigure}{.49\textwidth}
    \includegraphics[width=\textwidth]{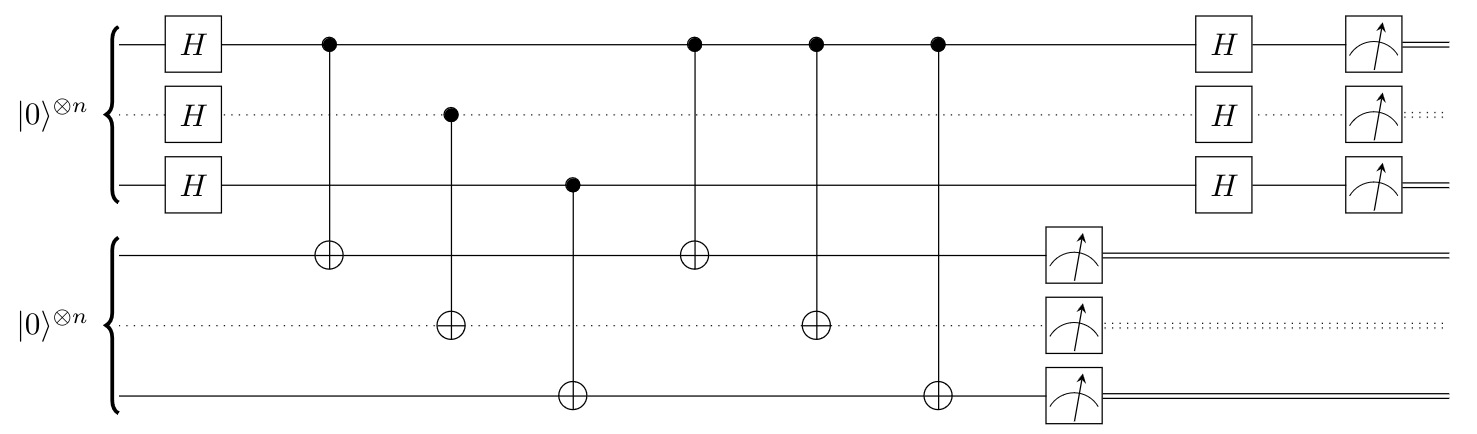}
    \caption{\emph{Maximal} number of two-qubit gates.}
    \end{subfigure}
    \hfill
     \begin{subfigure}{.49\textwidth}
    \includegraphics[width=\textwidth]{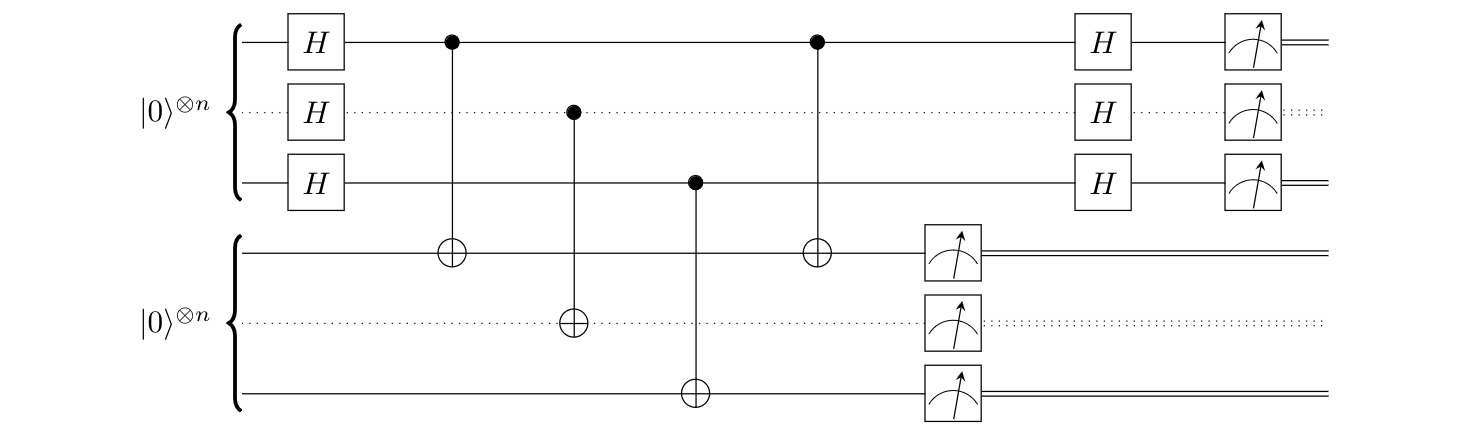}
    \caption{\emph{Minimal} number of two-qubit gates.}
    \end{subfigure}
    \caption{
        Complex and simple oracle for Simon's algorithm.
        Each dotted wire represents $(n-2)$ qubits, and all operations on dotted wires are repeated following the pattern across all qubits.
    }
    \label{fig:oracle}
\end{figure}

However, to implement the algorithm on real hardware, $U_f$, clearly does need to specified.
On NISQ devices every single gate operation is accompanied by its native noisiness, and hence one would expect the performance of the entire algorithm to be dependent on the explicit choice of $U_f$.
Luckily, Simon's algorithm is simple enough that one can easily identify the ``worst and best case scenarios.''
The worst case scenario corresponds to the $U_f$ with the \emph{maximal} number of two-qubit operations and the best case scenario to the \emph{minimal} number, see Figs.~\ref{fig:oracle}.
For brevity, we refer to these extreme cases as ``complex'' and ``simple'' oracles.

More explicitly, the algorithm is implement as follows:
First, we defined a function $f$ on bitstrings of size $n$, where $s$ was the string of $n$ 1s.
Then we constructed the corresponding oracle $U_f$ for the complex and simple cases.
Next, we allocated two registers of $n$ qubits on the device.
As depicted in Fig.~\ref{fig:oracle}, we applied a Hadamard transform to the first register, then the function $U_f$ to the first and second register, and then a second Hadamard transform to the first register.
The algorithm is concluded by measuring all qubits to obtain an output bitstring.
We extracted the bitstring corresponding to the first register, and then computed its product with $s$.
Finally, we repeated the algorithm for many shots, recording the percentage of shots that yielded invalid measurement results.
To quantify the performance of the algorithm, we define the \emph{algorithmic error rate}, which is percentage of results parallel to $s$.

\subsection{Results on IBM}

Many jobs on the real IBM quantum devices, Brisbane, Osaka, and Kyoto, were submitted through the Composer tool of IBM's Quantum web application\footnote{\url{https://quantum.ibm.com/composer}}.
This tool provides a graphical user interface through which to create and submit quantum circuit diagrams.
It also automatically transpiles algorithms written agnostic of hardware into a format that runs on a particular backend.
Transpiliation is a heuristic process that utilizes the current calibration of the machine \cite{ibm_transpilier}, which means that the gate operations utilized and physical qubits selected often varies between submissions of the algorithm.

We repeated this process three times and averaged the results.
For our purposes, $n$ ranged from 2 to 12, which meant a total of 33 jobs were executed on each device.
As alluded to above, we observed that the algorithm was transpiled differently for different runs of the same instance.
For example, our second run of the $n=12$ complex oracle utilized qubits 5-9 on IBM Osaka, while our third run of the same algorithm on the same device did not. We chose to allow for this variation in the physical implementation of the algorithm as it allowed the IBM system to dynamically optimize for the current characterization of the device.
Hence, all of our experiments are as near as possible to the most efficient implementation of the algorithm on each device at the time of execution. 

We also conducted experiments on the noisy simulators provided for the devices.
To this end, we utilized the IBM AER local simulator, seeded using error models derived from the real devices.
As per the Qiskit documentation, this noise model consists of single-qubit gate error, two-qubit gate errors, and the single qubit readout errors \cite{ibm_documentation}.
To generate the data for these noisy simulators we followed the same process as described above, with the following two modifications:
first, we repeated the entire process thirty times instead of three; second, we obtained the latest noise characteristic calibration data from the IBM API interface between each iteration of the experiment (i.e. the same model was used for $n \in [2,12]$, but a new model was recomputed before $n$ reset).

\begin{figure}
    \centering
    \includegraphics[width=\figsize\linewidth]{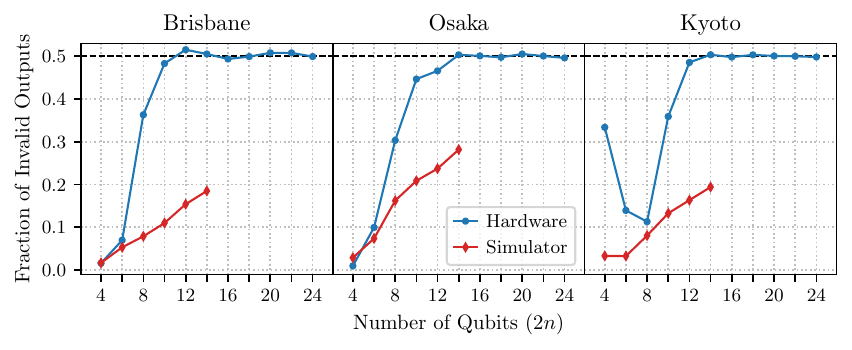}
    \caption{
        Simon's algorithm performed for the complex oracle at various sizes on IBM devices and simulators, cf. Fig.~\ref{fig:oracle}(a).
        By $\sim$12 qubits the real device hovers around $50\%$ algorithmic error, which is indistinguishable from randomly guessing solutions to the problem from the space of all possibilities.
    }
    \label{fig:ibm-comparison}
\end{figure}

\begin{figure}
    \centering
    \includegraphics[width=\textwidth]{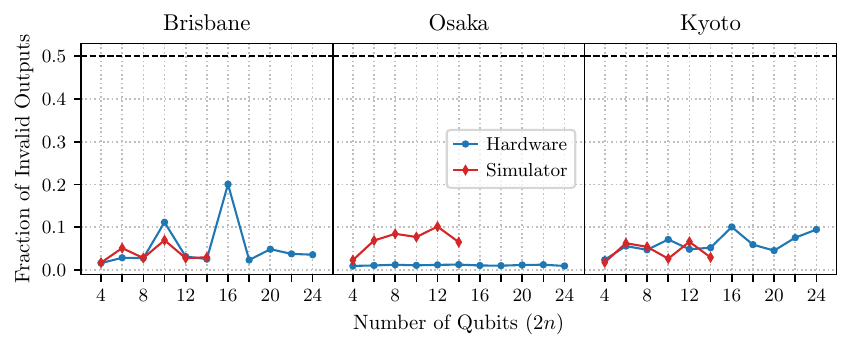}
    \caption{
        Simon's algorithm performed for the simple oracle at various sizes on IBM devices and simulators, cf. Fig.~\ref{fig:oracle}(b).
    }
    \label{fig:ibm-osaka-simple}
\end{figure}

Figure~\ref{fig:ibm-comparison} shows our results for the complex oracle requiring the maximal number of two-qubit gates, and in Fig.~\ref{fig:ibm-osaka-simple} we collect the results for the simplest oracle.
We observe that for already moderate sizes of the problem, $n>8$, the more complicated oracle leads to a failure of the algorithm.
Interesting, the simplest oracle performs remarkably well for all problem sizes.

More curious is the output of the noisy simulators.
While for the simpler case in Fig.~\ref{fig:ibm-osaka-simple} the output of the simulator and the hardware are qualitatively consistent, this is not the case for the more complicated oracle.
In fact, in this case the AER local simulators predict a roughly linear scaling in algorithmic error as the problem size increases for all devices.

A plausible hypothesis is that this increase in error coincides with the addition of a number of swap gates into the algorithm, as the chip layout cannot provide all the needed direct connections for problems of this size.
Interestingly, however, the simulated algorithm also inserts swap gates of the same pattern, but the jump in error is not observed.
Thus, the dramatic jump in algorithmic error observed on the hardware may be due to correlated errors that accompany the addition of the swap operations which is not captured in the simulator's noise model. 

\begin{figure}
    \centering
        \centering
        \includegraphics[width=.3\textwidth]{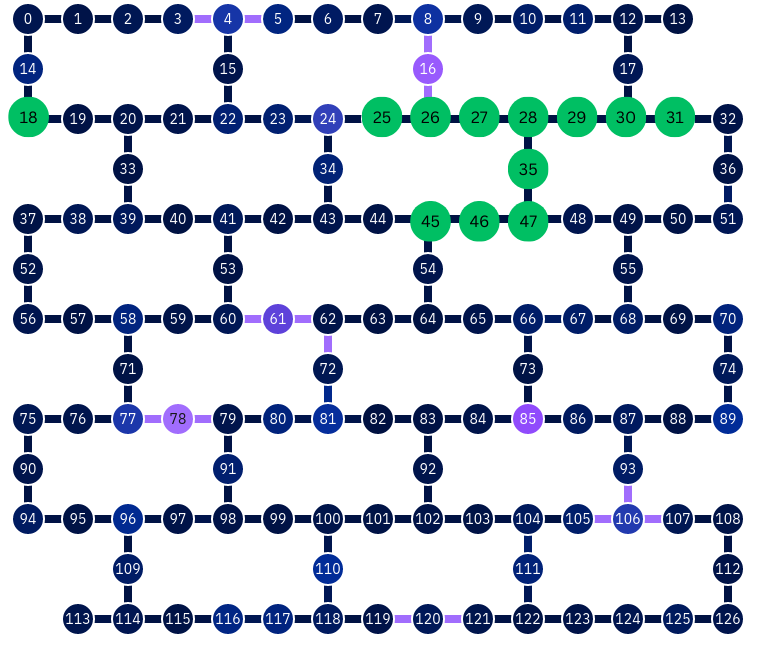}
        \label{fig:layout5}
    \caption{
        Example of a transpiled circuit:
        Tanspiled Simon's algorithm of the complex oracle for $n=5$ on IBM Osaka.
        Active qubits are marked in green.
        Note that after transpiliation qubit 18, although active, is idle for the duration of the algorithm and does not interact with the other active qubits.
        All other active qubits interact directly with one or two other qubits via entangling operations.
    }
    \label{fig:circuitToLayout}
\end{figure}

This hypothesis is further corroborated by inspecting the transpiled quantum circuits.
In Fig.~\ref{fig:circuitToLayout} we depict one example for $n=5$ on IBM Osaka, in which we have marked the ``active'' qubits in the QPU.
We notice that spatially separated qubits are active, which necessitates the physical implementation of two-qubit gates across substantial portions of the QPU.
The natural question arises to what degree the two-qubit gate error increases with the spatial separation between the qubits.

\begin{figure}
    \centering
    \includegraphics[width=\figsize\linewidth]{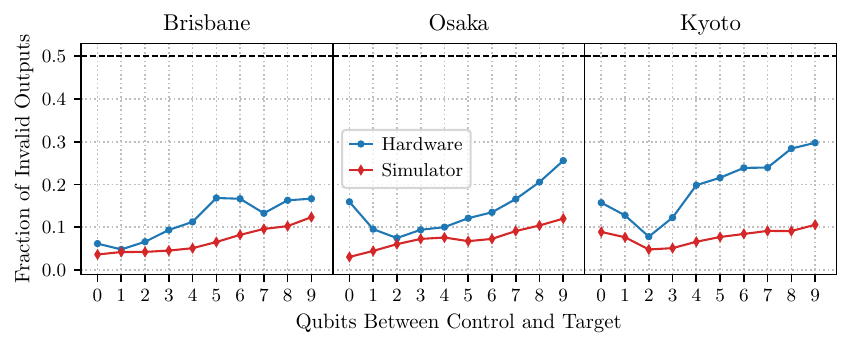}
    \caption{
        Failure of CNOT gates as a function of spatial separation.
        Prior to compilation the control bit was selected to be qubit 39 for each experiment, and the target bit ranged from bit 40 to 49.
    }
    \label{fig:ibm-cnot}
\end{figure}

In Fig.~\ref{fig:ibm-cnot} we collect our findings for the error rates of the CNOT gate as a function of the spatial separation of the control and target qubits.
As expected the error rate grows as a function of distance, but we also observe that the noisy simulators underestimate this effect.
While it cannot be excluded that there are further hardware and software issues that contribute to the failure of Simon's algorithm with the most complicated oracle, we can conclude that two-qubit operations between spatially separated qubits are a significant source of error.

Finally, we reiterate that after about 10 qubits the algorithmic error observed for the complex oracle on the real hardware is approximately $50\%$.
This indicates a complete failure of Simon's algorithm.
Randomly outputting bitstrings sampled uniformly from the space of all bitstrings of size $n$ would yield an error rate of approximately $50\%$, which is indistinguishable from the results observed here.

\subsection{Results on IonQ}

Since the tapped ion NISQ device by IonQ offers all-to-all connectivity, we only studied the complex oracle.
The jobs were submitted via the IonQ APIs using a hardware-agnostic Qiskit program that was transpilied for each IonQ device.
Again this means that the actual algorithm physically implemented may have varied for each job performed.
For IonQ Aria 1 we conducting two iterations of experiments for $n\in[2,12]$, with 8096 shots per job.
For IonQ Forte we only conducted the process once with 4096 shots per value of $n$, with $n$ ranging from 2 to 17\footnote{This simplification was mandated by budget and time constraints. At the time of publication Forte is only available through one hour reservations costing \$7000.}.
The corresponding 16 jobs were written in Qiskit, transpiled for Forte, and then submitted to the device.

In complete analogy to the experiments on IBM's platforms, we made use of the cloud-based noisy simulator provided for Aria (currently there is no noisy simulator offered for Forte).
As before, we performed 30 repetitions of our experiment on this simulator.
To the best of our knowledge, IonQ does not provide information about how often this simulator is recalibrated, nor what noise characteristics are used in its construction.
However, it does state that the results of the simulator are only broadly similar to the results of the physical device \cite{ionq_documentation}.

We also performed 30 repetitions of our experiment on the IonQ Harmony simulator.
This device has 11 qubits, which allowed us to perform runs with $n\in[2,5]$.
Unfortunately the real Harmony device was offline during the duration of our study, so we were not able to generate data on this computer.

\begin{figure}
    \centering
    \includegraphics[width=\figsize\linewidth]{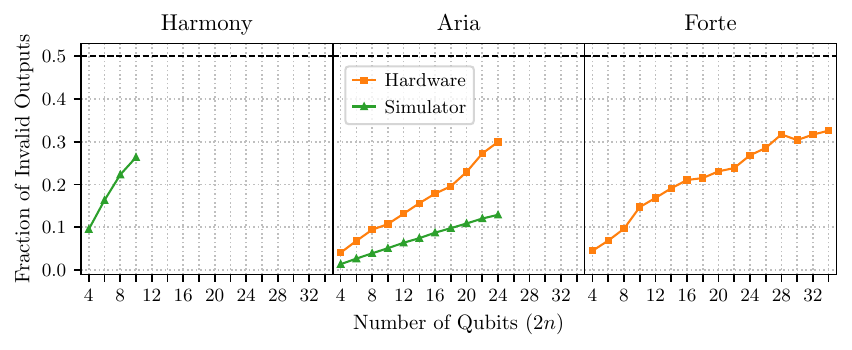}
    \caption{
        Simon's algorithm performed at various sizes on IonQ devices and simulators.
        Note that the Forte device does not have a corresponding simulator, and the real Harmony device was offline at the time of experimentation.
    }
    \label{fig:ionq-comparison}
\end{figure}

In Fig.~\ref{fig:ionq-comparison} we show our results.
While IonQ is not explicit about the details of their noise models, we see from Fig.~\ref{fig:ionq-comparison} that the Aria simulator predicts a roughly linear scaling.
This matches the general behavior of the real Aria device, however we observe that the simulator consistently under-approximates the observed error by a factor of about 2.
The Forte device also exhibits a roughly linear scaling pattern, similar to the Aria device.
This indicates two things: first, that the IonQ devices scale consistently throughout the ``intermediate scale,'' and second, IonQ simulators agree with the general behavior of the real devices (up to a constant factor).

Interestingly, the behavior of the algorithmic error rate for the devices of IBM and IonQ are markedly different.
However, for the complex oracle neither of the NISQ platforms appears capable to support real quantum advantage.
Assuming that quantum advantage would be present for $\sim$ 53 qubits, and extrapolating the error rates to such large systems, our results suggest that no studied devices would have an error rate of less than 50\% at this scale.

\section{Concluding remarks}
\label{conclusion}

In conclusion, we performed an benchmarking analysis of the algorithmic error of six NISQ devices using two implementations of Simon's algorithm.
For the most complex oracle, we found that all noisy simulators predict the algorithmic error to scale linearly with the problem size.
This prediction was corroborated by the results obtained on IonQ's trapped ion NISQ devices.
For IBM's superconduting QPUs we found that the choice of the oracle function makes an enormous difference in the performance of the algorithm.
One crucial source of error was traced back to the sensitivity of CNOT gate operations to the spatial separation of qubits on the hardware.

Our findings have unveiled and re-emphasized several facts about NISQ devices.
QPUs based on trapped ions might have some computational advantages due to their all-to-all connectivity.
For these devices, less attention needs to be paid to how an algorithm is transpiled onto the hardware.
Their obvious shortcoming is still the limited size.

The obvious advantage of superconducting devices is their size and scalability.
However, we clearly demonstrated that in the design of algorithms, or rather their transpilation, special attention needs to be paid to the topology of the QPU.
For instance, two-qubit operations on spatially separated qubits need to be avoided.

Our work is only a first step of a larger research agenda.
As we discussed above, Simon's algorithm is just one representative of hidden subgroup problems.
Other examples include the Deutsch-Jozsa algorithm, the Bernstein-Vazirani algorithm, and Shor's algorithm \cite{bible}.
The obvious question arises whether our findings are specific to Simon's, or whether one would expect similar performance of the other algorithms.
Interestingly, all of these algorithms assume access to noise-free qubits, and studying their behavior on NISQ devices would build a more developed picture of the current landscape of algorithmic error.

Even more importantly, such a complete picture would allow a better understanding for when to expect quantum advantage.
In fact, one should investigate the amount of error that Simon's algorithm can tolerate without loss of quantum advantage.
The computational advantage of Simon's algorithm is predicated on the ability to construct a system of $n-1$ linearly independent values orthogonal to the ``secret string'' $s$.
Such a system can be constructed with noise-free qubits in linear time, and then solved classically in polynomial-time.
With noisy qubits, however, the resulting system of equations will have some equations that are \textit{not} orthogonal to $s$.
Hence the problem becomes solving a noisy system of boolean linear equations.
Some effort has been devoted into algorithms for maximally satisfying such systems \cite{soft_solutions}, and a computational complexity result suggesting an exponential scaling for this problem has been derived \cite{noisy_systems}.
Further work is required to establish an exponential scaling for these noisy boolean linear systems, and then to determine at what value of algorithmic error the noise overwhelms the advantage of the quantum computation.

\section*{Availability of data and materials}

All code and data for this project can be accessed at a dedicated GitHub repository \url{https://github.com/reecejrobertson/simons-algorithm}.

\section*{Acknowledgments}

S.D. acknowledges support from the John Templeton Foundation under Grant No. 62422.
IBM Quantum services were used for this work.
The views expressed are those of the authors and do not reflect the official policy or position of IBM or the IBM Quantum team.

\appendix

\section{Physical parameters of the NISQ devices}

In Tab.~\ref{tab:comparison} we collect the physical parameters of the studied NISQ devices.

\begin{table}[htbp]
    \centering
    \begin{tabular}{|c|c|c|c|c|c|c|}
        \hline
        Parameter & Brisbane & Osaka & Kyoto & Forte & Aria 1 & Harmony \\
        \hline
        \hline
        T1 Time & 213.12 $\mu$s & 297.17 $\mu$s & 215.43 $\mu$s & 100 s & 100 s & 10000 s \\
        \hline
        T2 Time & 145.97 $\mu$s & 127.23 $\mu$s & 109.44 $\mu$s & 1 s & 1 s & 0.2 s \\
        \hline
        2-Qubit Gate Speed & 660 ns & 660 ns & 660 ns & 970 $\mu$s & 600 $\mu$s & 200 $\mu$s \\
        \hline
        1-Qubit Gate Error (\%) & 0.03 & 0.03 & 0.03 & 0.09 & 0.06 & 0.67 \\
        \hline
        2-Qubit Gate Error (\%) & 0.74 & 0.93 & 0.92 & 0.74 & 8.57 & 3.07 \\
        \hline
        Average Readout Error (\%) & 1.32 & 2.18 & 1.48 & 0.5 & 0.52 & 0.42 \\
        \hline
        Topology & Eagle r3 & Eagle r3 & Eagle r3 & all-to-all & all-to-all & all-to-all \\
        \hline
        Native Gates & IBM & IBM & IBM & IonQ & IonQ & IonQ \\
        \hline
    \end{tabular}
    \caption{
        Performance metrics for IBM and IonQ devices.
        Note that for one and two qubit gate errors and speeds, IBM reports median values while IonQ reports average values.
        Additionally, IBM native gate set includes: ECR, ID, RZ, SX, and X, while the IonQ native gate set consists of: MS, GPI, and GPI2.
        Retrieved on June 11, 2024 from \url{https://quantum.ibm.com/services/resources} and \url{https://cloud.ionq.com/backends}.
    }
    \label{tab:comparison}
\end{table}

\section{Comparison of the IBM devices before and after update}

\begin{figure}
    \centering
    \includegraphics[width=.45\textwidth]{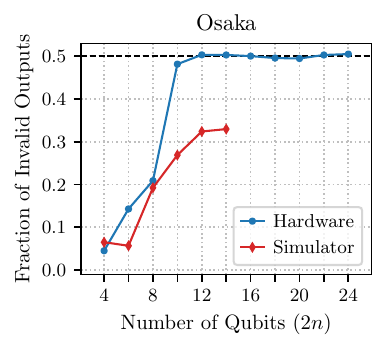}
    \caption{
        Simon's algorithm performed for the complex oracle at various sizes on IBM Osaka and simulator, cf. Fig.~\ref{fig:oracle}(a), after the major update to Qiskit.
    }
    \label{fig:ibm-osaka-post-update}
\end{figure}

A rerun of the experiment discussed above following the major Qiskit and IBM Quantum update of early May 2024.
In Fig.~\ref{fig:ibm-osaka-post-update} we show our findings.
We see that the performance of both the simulator and real device is comparable to what is shown in Figure \ref{fig:ibm-comparison}, with an obvious improvement in the 2-4 qubit range here.
This change brings Osaka more inline with the other devices tested, and validates that the poor performance originally observed was likely due to an unlucky run on the device, rather than reflecting a fault in the hardware.
Additionally, it is worth noting that after the update all jobs submitted to Osaka completed after 5-6 seconds of runtime, as opposed to the ~20 second runtime originally observed.

\bibliographystyle{elsarticle-num}
\bibliography{main}

\providecommand{\noopsort}[1]{}\providecommand{\singleletter}[1]{#1}%
\begin{thebibliography}{10}
\expandafter\ifx\csname url\endcsname\relax
  \def\url#1{\texttt{#1}}\fi
\expandafter\ifx\csname urlprefix\endcsname\relax\def\urlprefix{URL }\fi
\expandafter\ifx\csname href\endcsname\relax
  \def\href#1#2{#2} \def\path#1{#1}\fi

\bibitem{invest}
McKinsey, \href{https://www.mckinsey.com/capabilities/mckinsey-digital/our-insights/quantum-technology-sees-record-investments-progress-on-talent-gap}{Quantum technology sees record investments, progress on talent gap}.
\newline\urlprefix\url{https://www.mckinsey.com/capabilities/mckinsey-digital/our-insights/quantum-technology-sees-record-investments-progress-on-talent-gap}

\bibitem{bible}
M.~A. Nielsen, I.~L. Chuang, Quantum Computation and Quantum Information: 10th Anniversary Edition, Cambridge University Press, 2010.

\bibitem{Sanders2017}
B.~C. Sanders, \href{http://dx.doi.org/10.1088/978-0-7503-1536-4}{How to Build a Quantum Computer}, 2399-2891, IOP Publishing, 2017.
\newblock \href {https://doi.org/10.1088/978-0-7503-1536-4} {\path{doi:10.1088/978-0-7503-1536-4}}.
\newline\urlprefix\url{http://dx.doi.org/10.1088/978-0-7503-1536-4}

\bibitem{Savage2017}
N.~Savage, \href{https://www.scientificamerican.com/article/quantum-computers-compete-for-supremacy/}{Quantum computers compete for supremacy}, Scientific American July (2017).
\newline\urlprefix\url{https://www.scientificamerican.com/article/quantum-computers-compete-for-supremacy/}

\bibitem{Arute2019}
F.~Arute, et~al., \href{https://doi.org/10.1038/s41586-019-1666-5}{Quantum supremacy using a programmable superconducting processor}, Nature 574~(7779) (2019) 505--510.
\newblock \href {https://doi.org/10.1038/s41586-019-1666-5} {\path{doi:10.1038/s41586-019-1666-5}}.
\newline\urlprefix\url{https://doi.org/10.1038/s41586-019-1666-5}

\bibitem{HanSen2020Science}
H.-S. Zhong, H.~Wang, Y.-H. Deng, M.-C. Chen, L.-C. Peng, Y.-H. Luo, J.~Qin, D.~Wu, X.~Ding, Y.~Hu, P.~Hu, X.-Y. Yang, W.-J. Zhang, H.~Li, Y.~Li, X.~Jiang, L.~Gan, G.~Yang, L.~You, Z.~Wang, L.~Li, N.-L. Liu, C.-Y. Lu, J.-W. Pan, \href{https://www.science.org/doi/abs/10.1126/science.abe8770}{Quantum computational advantage using photons}, Science 370~(6523) (2020) 1460--1463.
\newblock \href {http://arxiv.org/abs/https://www.science.org/doi/pdf/10.1126/science.abe8770} {\path{arXiv:https://www.science.org/doi/pdf/10.1126/science.abe8770}}, \href {https://doi.org/10.1126/science.abe8770} {\path{doi:10.1126/science.abe8770}}.
\newline\urlprefix\url{https://www.science.org/doi/abs/10.1126/science.abe8770}

\bibitem{Wu2021PRL}
Y.~Wu, W.-S. Bao, S.~Cao, F.~Chen, M.-C. Chen, X.~Chen, T.-H. Chung, H.~Deng, Y.~Du, D.~Fan, M.~Gong, C.~Guo, C.~Guo, S.~Guo, L.~Han, L.~Hong, H.-L. Huang, Y.-H. Huo, L.~Li, N.~Li, S.~Li, Y.~Li, F.~Liang, C.~Lin, J.~Lin, H.~Qian, D.~Qiao, H.~Rong, H.~Su, L.~Sun, L.~Wang, S.~Wang, D.~Wu, Y.~Xu, K.~Yan, W.~Yang, Y.~Yang, Y.~Ye, J.~Yin, C.~Ying, J.~Yu, C.~Zha, C.~Zhang, H.~Zhang, K.~Zhang, Y.~Zhang, H.~Zhao, Y.~Zhao, L.~Zhou, Q.~Zhu, C.-Y. Lu, C.-Z. Peng, X.~Zhu, J.-W. Pan, \href{https://link.aps.org/doi/10.1103/PhysRevLett.127.180501}{Strong quantum computational advantage using a superconducting quantum processor}, Phys. Rev. Lett. 127 (2021) 180501.
\newblock \href {https://doi.org/10.1103/PhysRevLett.127.180501} {\path{doi:10.1103/PhysRevLett.127.180501}}.
\newline\urlprefix\url{https://link.aps.org/doi/10.1103/PhysRevLett.127.180501}

\bibitem{Zhong2021PRL}
H.-S. Zhong, Y.-H. Deng, J.~Qin, H.~Wang, M.-C. Chen, L.-C. Peng, Y.-H. Luo, D.~Wu, S.-Q. Gong, H.~Su, Y.~Hu, P.~Hu, X.-Y. Yang, W.-J. Zhang, H.~Li, Y.~Li, X.~Jiang, L.~Gan, G.~Yang, L.~You, Z.~Wang, L.~Li, N.-L. Liu, J.~J. Renema, C.-Y. Lu, J.-W. Pan, \href{https://link.aps.org/doi/10.1103/PhysRevLett.127.180502}{Phase-programmable gaussian boson sampling using stimulated squeezed light}, Phys. Rev. Lett. 127 (2021) 180502.
\newblock \href {https://doi.org/10.1103/PhysRevLett.127.180502} {\path{doi:10.1103/PhysRevLett.127.180502}}.
\newline\urlprefix\url{https://link.aps.org/doi/10.1103/PhysRevLett.127.180502}

\bibitem{Madsen2022Nature}
L.~S. Madsen, F.~Laudenbach, M.~F. Askarani, F.~Rortais, T.~Vincent, J.~F.~F. Bulmer, F.~M. Miatto, L.~Neuhaus, L.~G. Helt, M.~J. Collins, A.~E. Lita, T.~Gerrits, S.~W. Nam, V.~D. Vaidya, M.~Menotti, I.~Dhand, Z.~Vernon, N.~Quesada, J.~Lavoie, \href{https://doi.org/10.1038/s41586-022-04725-x}{Quantum computational advantage with a programmable photonic processor}, Nature 606~(7912) (2022) 75--81.
\newblock \href {https://doi.org/10.1038/s41586-022-04725-x} {\path{doi:10.1038/s41586-022-04725-x}}.
\newline\urlprefix\url{https://doi.org/10.1038/s41586-022-04725-x}

\bibitem{Zhu2022SB}
Q.~Zhu, S.~Cao, F.~Chen, M.-C. Chen, X.~Chen, T.-H. Chung, H.~Deng, Y.~Du, D.~Fan, M.~Gong, C.~Guo, C.~Guo, S.~Guo, L.~Han, L.~Hong, H.-L. Huang, Y.-H. Huo, L.~Li, N.~Li, S.~Li, Y.~Li, F.~Liang, C.~Lin, J.~Lin, H.~Qian, D.~Qiao, H.~Rong, H.~Su, L.~Sun, L.~Wang, S.~Wang, D.~Wu, Y.~Wu, Y.~Xu, K.~Yan, W.~Yang, Y.~Yang, Y.~Ye, J.~Yin, C.~Ying, J.~Yu, C.~Zha, C.~Zhang, H.~Zhang, K.~Zhang, Y.~Zhang, H.~Zhao, Y.~Zhao, L.~Zhou, C.-Y. Lu, C.-Z. Peng, X.~Zhu, J.-W. Pan, \href{https://www.sciencedirect.com/science/article/pii/S2095927321006733}{Quantum computational advantage via 60-qubit 24-cycle random circuit sampling}, Science Bulletin 67~(3) (2022) 240--245.
\newblock \href {https://doi.org/https://doi.org/10.1016/j.scib.2021.10.017} {\path{doi:https://doi.org/10.1016/j.scib.2021.10.017}}.
\newline\urlprefix\url{https://www.sciencedirect.com/science/article/pii/S2095927321006733}

\bibitem{King2024arXiv}
A.~D. King, A.~Nocera, M.~M. Rams, J.~Dziarmaga, R.~Wiersema, W.~Bernoudy, J.~Raymond, N.~Kaushal, N.~Heinsdorf, R.~Harris, K.~Boothby, F.~Altomare, A.~J. Berkley, M.~Boschnak, K.~Chern, H.~Christiani, S.~Cibere, J.~Connor, M.~H. Dehn, R.~Deshpande, S.~Ejtemaee, P.~Farré, K.~Hamer, E.~Hoskinson, S.~Huang, M.~W. Johnson, S.~Kortas, E.~Ladizinsky, T.~Lai, T.~Lanting, R.~Li, A.~J.~R. MacDonald, G.~Marsden, C.~C. McGeoch, R.~Molavi, R.~Neufeld, M.~Norouzpour, T.~Oh, J.~Pasvolsky, P.~Poitras, G.~Poulin-Lamarre, T.~Prescott, M.~Reis, C.~Rich, M.~Samani, B.~Sheldan, A.~Smirnov, E.~Sterpka, B.~T. Clavera, N.~Tsai, M.~Volkmann, A.~Whiticar, J.~D. Whittaker, W.~Wilkinson, J.~Yao, T.~J. Yi, A.~W. Sandvik, G.~Alvarez, R.~G. Melko, J.~Carrasquilla, M.~Franz, M.~H. Amin, Computational supremacy in quantum simulation, arXiv preprint arXiv:2403.00910 (2024).

\bibitem{resilient}
M.~P. da~Silva, C.~Ryan-Anderson, J.~M. Bello-Rivas, A.~Chernoguzov, J.~M. Dreiling, C.~Foltz, F.~Frachon, J.~P. Gaebler, T.~M. Gatterman, L.~Grans-Samuelsson, D.~Hayes, N.~Hewitt, J.~Johansen, D.~Lucchetti, M.~Mills, S.~A. Moses, B.~Neyenhuis, A.~Paz, J.~Pino, P.~Siegfried, J.~Strabley, A.~Sundaram, D.~Tom, S.~J. Wernli, M.~Zanner, R.~P. Stutz, K.~M. Svore, Demonstration of logical qubits and repeated error correction with better-than-physical error rates (2024).
\newblock \href {http://arxiv.org/abs/2404.02280} {\path{arXiv:2404.02280}}.

\bibitem{Preskill2018quantum}
J.~Preskill, \href{https://doi.org/10.22331/q-2018-08-06-79}{Quantum {C}omputing in the {NISQ} era and beyond}, {Quantum} 2 (2018) 79.
\newblock \href {https://doi.org/10.22331/q-2018-08-06-79} {\path{doi:10.22331/q-2018-08-06-79}}.
\newline\urlprefix\url{https://doi.org/10.22331/q-2018-08-06-79}

\bibitem{taxonomy}
S.~S. Gill, A.~Kumar, H.~Singh, M.~Singh, K.~Kaur, M.~Usman, R.~Buyya, \href{https://onlinelibrary.wiley.com/doi/abs/10.1002/spe.3039}{Quantum computing: A taxonomy, systematic review and future directions}, Software: Practice and Experience 52~(1) (2022) 66--114.
\newblock \href {http://arxiv.org/abs/https://onlinelibrary.wiley.com/doi/pdf/10.1002/spe.3039} {\path{arXiv:https://onlinelibrary.wiley.com/doi/pdf/10.1002/spe.3039}}, \href {https://doi.org/https://doi.org/10.1002/spe.3039} {\path{doi:https://doi.org/10.1002/spe.3039}}.
\newline\urlprefix\url{https://onlinelibrary.wiley.com/doi/abs/10.1002/spe.3039}

\bibitem{nisq_computing}
J.~W.~Z. Lau, K.~H. Lim, H.~Shrotriya, L.~C. Kwek, \href{https://doi.org/10.1007/s43673-022-00058-z}{Nisq computing: where are we and where do we go?}, AAPPS Bulletin 32~(1) (2022) 27.
\newblock \href {https://doi.org/10.1007/s43673-022-00058-z} {\path{doi:10.1007/s43673-022-00058-z}}.
\newline\urlprefix\url{https://doi.org/10.1007/s43673-022-00058-z}

\bibitem{survey}
Z.~Yang, M.~Zolanvari, R.~Jain, A survey of important issues in quantum computing and communications, IEEE Communications Surveys \& Tutorials 25~(2) (2023) 1059--1094.
\newblock \href {https://doi.org/10.1109/COMST.2023.3254481} {\path{doi:10.1109/COMST.2023.3254481}}.

\bibitem{evaluation}
T.~Patel, A.~Potharaju, B.~Li, R.~B. Roy, D.~Tiwari, Experimental evaluation of nisq quantum computers: Error measurement, characterization, and implications, in: SC20: International Conference for High Performance Computing, Networking, Storage and Analysis, 2020, pp. 1--15.
\newblock \href {https://doi.org/10.1109/SC41405.2020.00050} {\path{doi:10.1109/SC41405.2020.00050}}.

\bibitem{benchmarking}
M.~Kordzanganeh, M.~Buchberger, B.~Kyriacou, M.~Povolotskii, W.~Fischer, A.~Kurkin, W.~Somogyi, A.~Sagingalieva, M.~Pflitsch, A.~Melnikov, \href{https://onlinelibrary.wiley.com/doi/abs/10.1002/qute.202300043}{Benchmarking simulated and physical quantum processing units using quantum and hybrid algorithms}, Advanced Quantum Technologies 6~(8) (2023) 2300043.
\newblock \href {http://arxiv.org/abs/https://onlinelibrary.wiley.com/doi/pdf/10.1002/qute.202300043} {\path{arXiv:https://onlinelibrary.wiley.com/doi/pdf/10.1002/qute.202300043}}, \href {https://doi.org/https://doi.org/10.1002/qute.202300043} {\path{doi:https://doi.org/10.1002/qute.202300043}}.
\newline\urlprefix\url{https://onlinelibrary.wiley.com/doi/abs/10.1002/qute.202300043}

\bibitem{stability}
S.~Dasgupta, T.~S. Humble, Stability of noisy quantum computing devices (2021).
\newblock \href {http://arxiv.org/abs/2105.09472} {\path{arXiv:2105.09472}}.

\bibitem{characterizing}
S.~Dasgupta, T.~S. Humble, Characterizing the stability of nisq devices, in: 2020 IEEE International Conference on Quantum Computing and Engineering (QCE), 2020, pp. 419--429.
\newblock \href {https://doi.org/10.1109/QCE49297.2020.00059} {\path{doi:10.1109/QCE49297.2020.00059}}.

\bibitem{gate-based}
A.~Cornelissen, J.~Bausch, A.~Gilyén, Scalable benchmarks for gate-based quantum computers (2021).
\newblock \href {http://arxiv.org/abs/2104.10698} {\path{arXiv:2104.10698}}.

\bibitem{supermarq}
T.~Tomesh, P.~Gokhale, V.~Omole, G.~S. Ravi, K.~N. Smith, J.~Viszlai, X.-C. Wu, N.~Hardavellas, M.~R. Martonosi, F.~T. Chong, Supermarq: A scalable quantum benchmark suite, in: 2022 IEEE International Symposium on High-Performance Computer Architecture (HPCA), 2022, pp. 587--603.
\newblock \href {https://doi.org/10.1109/HPCA53966.2022.00050} {\path{doi:10.1109/HPCA53966.2022.00050}}.

\bibitem{qpack}
H.~Donkers, K.~Mesman, Z.~Al-Ars, M.~Möller, Qpack scores: Quantitative performance metrics for application-oriented quantum computer benchmarking (2022).
\newblock \href {http://arxiv.org/abs/2205.12142} {\path{arXiv:2205.12142}}.

\bibitem{qasmbench}
A.~Li, S.~Stein, S.~Krishnamoorthy, J.~Ang, \href{https://doi.org/10.1145/3550488}{Qasmbench: A low-level quantum benchmark suite for nisq evaluation and simulation}, ACM Transactions on Quantum Computing 4~(2) (feb 2023).
\newblock \href {https://doi.org/10.1145/3550488} {\path{doi:10.1145/3550488}}.
\newline\urlprefix\url{https://doi.org/10.1145/3550488}

\bibitem{nisq_analyzer}
M.~Salm, J.~Barzen, U.~Breitenb{\"u}cher, F.~Leymann, B.~Weder, K.~Wild, The nisq analyzer: Automating the selection of quantum computers for quantum algorithms, in: S.~Dustdar (Ed.), Service-Oriented Computing, Springer International Publishing, Cham, 2020, pp. 66--85.

\bibitem{shor}
P.~Shor, Algorithms for quantum computation: discrete logarithms and factoring, in: Proceedings 35th Annual Symposium on Foundations of Computer Science, 1994, pp. 124--134.
\newblock \href {https://doi.org/10.1109/SFCS.1994.365700} {\path{doi:10.1109/SFCS.1994.365700}}.

\bibitem{chuang}
I.~L. Chuang, L.~M.~K. Vandersypen, X.~Zhou, D.~W. Leung, S.~Lloyd, \href{https://doi.org/10.1038/30181}{Experimental realization of a quantum algorithm}, Nature 393~(6681) (1998) 143--146.
\newblock \href {https://doi.org/10.1038/30181} {\path{doi:10.1038/30181}}.
\newline\urlprefix\url{https://doi.org/10.1038/30181}

\bibitem{frontier}
A.~Bouland, B.~Fefferman, Z.~Landau, Y.~Liu, Noise and the frontier of quantum supremacy, in: 2021 IEEE 62nd Annual Symposium on Foundations of Computer Science (FOCS), 2022, pp. 1308--1317.
\newblock \href {https://doi.org/10.1109/FOCS52979.2021.00127} {\path{doi:10.1109/FOCS52979.2021.00127}}.

\bibitem{nisq_complexity}
S.~Chen, J.~Cotler, H.-Y. Huang, J.~Li, \href{https://doi.org/10.1038/s41467-023-41217-6}{The complexity of nisq}, Nature Communications 14~(1) (2023) 6001.
\newblock \href {https://doi.org/10.1038/s41467-023-41217-6} {\path{doi:10.1038/s41467-023-41217-6}}.
\newline\urlprefix\url{https://doi.org/10.1038/s41467-023-41217-6}

\bibitem{sampling}
E.~Pelofske, J.~Golden, A.~Bärtschi, D.~O’Malley, S.~Eidenbenz, Sampling on nisq devices: "who’s the fairest one of all?", in: 2021 IEEE International Conference on Quantum Computing and Engineering (QCE), 2021, pp. 207--217.
\newblock \href {https://doi.org/10.1109/QCE52317.2021.00038} {\path{doi:10.1109/QCE52317.2021.00038}}.

\bibitem{simon}
D.~R. Simon, \href{https://doi.org/10.1137/S0097539796298637}{On the power of quantum computation}, SIAM Journal on Computing 26~(5) (1997) 1474--1483.
\newblock \href {http://arxiv.org/abs/https://doi.org/10.1137/S0097539796298637} {\path{arXiv:https://doi.org/10.1137/S0097539796298637}}, \href {https://doi.org/10.1137/S0097539796298637} {\path{doi:10.1137/S0097539796298637}}.
\newline\urlprefix\url{https://doi.org/10.1137/S0097539796298637}

\bibitem{strengths_weaknesses}
C.~H. Bennett, E.~Bernstein, G.~Brassard, U.~Vazirani, \href{https://doi.org/10.1137/S0097539796300933}{Strengths and weaknesses of quantum computing}, SIAM Journal on Computing 26~(5) (1997) 1510--1523.
\newblock \href {http://arxiv.org/abs/https://doi.org/10.1137/S0097539796300933} {\path{arXiv:https://doi.org/10.1137/S0097539796300933}}, \href {https://doi.org/10.1137/S0097539796300933} {\path{doi:10.1137/S0097539796300933}}.
\newline\urlprefix\url{https://doi.org/10.1137/S0097539796300933}

\bibitem{impact}
O.~Subasi, S.~Krishnamoorthy, The impact of logical errors on quantum algorithms (2023).
\newblock \href {http://arxiv.org/abs/2111.03733} {\path{arXiv:2111.03733}}.

\bibitem{Shor1994}
P.~Shor, Algorithms for quantum computation: discrete logarithms and factoring, in: Proceedings 35th Annual Symposium on Foundations of Computer Science, 1994, pp. 124--134.
\newblock \href {https://doi.org/10.1109/SFCS.1994.365700} {\path{doi:10.1109/SFCS.1994.365700}}.

\bibitem{mermin}
N.~D. Mermin, Quantum Computer Science: An Introduction, Cambridge University Press, 2007.

\bibitem{Cross2019PRA}
A.~W. Cross, L.~S. Bishop, S.~Sheldon, P.~D. Nation, J.~M. Gambetta, \href{https://link.aps.org/doi/10.1103/PhysRevA.100.032328}{Validating quantum computers using randomized model circuits}, Phys. Rev. A 100 (2019) 032328.
\newblock \href {https://doi.org/10.1103/PhysRevA.100.032328} {\path{doi:10.1103/PhysRevA.100.032328}}.
\newline\urlprefix\url{https://link.aps.org/doi/10.1103/PhysRevA.100.032328}

\bibitem{quantum_volume}
E.~Pelofske, A.~Bärtschi, S.~Eidenbenz, Quantum volume in practice: What users can expect from nisq devices, IEEE Transactions on Quantum Engineering 3 (2022) 1--19.
\newblock \href {https://doi.org/10.1109/TQE.2022.3184764} {\path{doi:10.1109/TQE.2022.3184764}}.

\bibitem{Koch2007PRA}
J.~Koch, T.~M. Yu, J.~Gambetta, A.~A. Houck, D.~I. Schuster, J.~Majer, A.~Blais, M.~H. Devoret, S.~M. Girvin, R.~J. Schoelkopf, \href{https://link.aps.org/doi/10.1103/PhysRevA.76.042319}{Charge-insensitive qubit design derived from the cooper pair box}, Phys. Rev. A 76 (2007) 042319.
\newblock \href {https://doi.org/10.1103/PhysRevA.76.042319} {\path{doi:10.1103/PhysRevA.76.042319}}.
\newline\urlprefix\url{https://link.aps.org/doi/10.1103/PhysRevA.76.042319}

\bibitem{Deffner2017Heliyon}
S.~Deffner, \href{https://www.sciencedirect.com/science/article/pii/S2405844017319369}{Demonstration of entanglement assisted invariance on ibm's quantum experience}, Heliyon 3~(11) (2017) e00444.
\newblock \href {https://doi.org/https://doi.org/10.1016/j.heliyon.2017.e00444} {\path{doi:https://doi.org/10.1016/j.heliyon.2017.e00444}}.
\newline\urlprefix\url{https://www.sciencedirect.com/science/article/pii/S2405844017319369}

\bibitem{Castelvecchi2023ibm}
D.~Castelvecchi, \href{https://www.nature.com/articles/d41586-023-03854-1}{Ibm releases first-ever 1,000-qubit quantum chip}, Nature 624~(7991) (2023) 238--238.
\newline\urlprefix\url{https://www.nature.com/articles/d41586-023-03854-1}

\bibitem{cloud}
G.~S. Ravi, K.~N. Smith, P.~Gokhale, F.~T. Chong, Quantum computing in the cloud: Analyzing job and machine characteristics, in: 2021 IEEE International Symposium on Workload Characterization (IISWC), 2021, pp. 39--50.
\newblock \href {https://doi.org/10.1109/IISWC53511.2021.00015} {\path{doi:10.1109/IISWC53511.2021.00015}}.

\bibitem{Allen2017IEEE}
S.~Allen, J.~Kim, D.~L. Moehring, C.~R. Monroe, Reconfigurable and programmable ion trap quantum computer, in: 2017 IEEE International Conference on Rebooting Computing (ICRC), 2017, pp. 1--3.
\newblock \href {https://doi.org/10.1109/ICRC.2017.8123665} {\path{doi:10.1109/ICRC.2017.8123665}}.

\bibitem{ibm_transpilier}
\href{https://docs.quantum.ibm.com/api/qiskit/transpiler}{Transpilier}.
\newline\urlprefix\url{https://docs.quantum.ibm.com/api/qiskit/transpiler}

\bibitem{ibm_documentation}
\href{https://qiskit.org/ecosystem/aer/tutorials/2_device_noise_simulation.html}{Device backend noise model simulations - {Qiskit} {Aer} 0.13.1}.
\newline\urlprefix\url{https://qiskit.org/ecosystem/aer/tutorials/2_device_noise_simulation.html}

\bibitem{ionq_documentation}
\href{https://ionq.com/docs/get-started-with-hardware-noise-model-simulation}{Get started with hardware noise model simulation}.
\newline\urlprefix\url{https://ionq.com/docs/get-started-with-hardware-noise-model-simulation}

\bibitem{soft_solutions}
T.~K. Moon, J.~O. Jensen, J.~H. Gunther, Soft solution of noisy linear gf(2) equations, in: 2022 Intermountain Engineering, Technology and Computing (IETC), 2022, pp. 1--6.
\newblock \href {https://doi.org/10.1109/IETC54973.2022.9796941} {\path{doi:10.1109/IETC54973.2022.9796941}}.

\bibitem{noisy_systems}
M.~Alekhnovich, More on average case vs approximation complexity, in: 44th Annual IEEE Symposium on Foundations of Computer Science, 2003. Proceedings., 2003, pp. 298--307.
\newblock \href {https://doi.org/10.1109/SFCS.2003.1238204} {\path{doi:10.1109/SFCS.2003.1238204}}.

\end{thebibliography}

\end{document}